\documentstyle[12pt,aaspp4]{article}
\def\hi   {\protect\ion{H}{1}}
\def\hii  {\protect\ion{H}{2}}
\def\sii  {\protect\ion{S}{2}}
\def\oi   {\protect\ion{O}{1}}
\received{6 May 1997}
\accepted{31 May 1997}
\lefthead{Eskridge \& White}
\righthead{NGC 6822 X-1}

\begin{document}

\title{The Nature of the X-Ray Point Source in the Bar of NGC 6822}

\author{Paul B.~Eskridge\footnotemark}
\footnotetext{paul@hera.astr.ua.edu}
\and
\author{Raymond E.~White III\footnotemark}
\footnotetext{white@merkin.astr.ua.edu}
\affil{Department of Physics and Astronomy, \break University of Alabama, 
Tuscaloosa, AL 35487}

\authoremail{paul@hera.astr.ua.edu, white@hera.astr.ua.edu}

\begin{abstract}
We have analysed archival {\it ROSAT} PSPC and {\it Einstein} HRI data for the
x-ray point source in the bar of NGC 6822.  The source decreased in x-ray flux
by at least half an order of magnitude in the 13 years between the HRI and PSPC
observations.  It has a PSPC flux of $f_X \approx 8 \times 10^{-14}~{\rm 
erg~cm^{-2}~s^{-1}}$, giving $L_X \approx 2.4 \times 10^{36}~{\rm erg~s^{-1}}$
for a distance of 500 kpc.  The source appears to be related to the optical 
emission-line object Ho 12.  It is unclear if Ho 12 is an \hii\ region or a 
supernova remnant, although the latter interpretation is better supported by 
the available optical data.  The x-rays are unlikely to be direct emission from 
a supernova remnant, due to the observed strong x-ray variability.  The x-ray 
spectrum of the source is very soft.  Acceptable fits are found for either a 
Raymond-Smith thermal model with kT$\approx$0.56 keV, or a blackbody model with 
kT$\approx$0.06 keV.  The latter model would place the source amongst the class 
of Super-Soft x-ray sources first found in the Magellanic Clouds.  If the 
thermal model is correct, the source appears similar to Galactic and Magellanic 
Cloud stellar-mass black hole binary candidates.  We have identified a list of 
potential optical counterparts from published photometry.  However, for either 
model discussed above, we expect the apparent magnitude of the optical 
counterpart to be in the range $21.5 \la m_V \la 24.0$.  Thus the optical 
counterpart may be below the limit of published photometry.
\end{abstract}

\section{Introduction}

NGC 6822 is the nearest dwarf irregular (dI) galaxy beyond the Galactic halo,
and has the distinction of being the first object clearly identified as an 
external stellar system (\markcite{p22}Perrine 1922; \markcite{h25}Hubble 
1925).  It has a Cepheid-based distance modulus of $(m-M)_{\circ} = 23.5$ 
(\markcite{m83}McAlary et al.~1983; \markcite{gav}Gallart, Aparicio \& 
V{\'\i}lchez 1996a), corresponding to a distance of $\sim$500 kpc.  Dwarf 
galaxies in the Local Group have much less rich x-ray source populations than 
do the more luminous galaxies, and they have therefore been relatively 
neglected.  This is the third in a series of papers that seek to redress this 
situation.  Previous papers in the series studied IC 1613 
(\markcite{e95}Eskridge 1995) and M32 (\markcite{ewd}Eskridge, White \& Davis 
1996).  

There have been a large number of studies of the stellar populations of NGC
6822, including global studies of the luminous stars (e.g.~\markcite{hea}Hodge 
et al.~1991 and references therein), and much deeper CCD studies of selected 
areas (e.g.~\markcite{w92}Wilson 1992a; \markcite{gav}Gallart et al.~1996a).  
All these studies show the young population quite clearly:  there is a strong 
blue main-sequence (MS), as well as yellow and red supergiants.  The existence 
of a substantial intermediate age population (from the presence of an extended 
asymptotic giant branch - AGB) is most obvious from the $VI$ photometry of 
\markcite{gav}Gallart et al.~(1996a).  While there is no unambiguous evidence 
for a true Pop II in the field of NGC 6822, the study of \markcite{gea}Gallart 
et al.~(1996b) strongly supports the existence of such a population.  There are 
$\sim$20 OB associations identified in NGC 6822 (\markcite{h77}Hodge 1977; 
\markcite{w92}Wilson 1992a).  \markcite{w92}Wilson (1992a) presents 
color-magnitude (CM) diagrams for 13 of these, down to $V \approx 22$.  These 
CM diagrams are only deep enough to show the OB main sequence stars and the 
supergiants.  Wilson finds a slope of the upper MS comparable to that found in 
IC 1613 (\markcite{f88}Freedman 1988), and other Local Group galaxies 
(e.g.~\markcite{f85}Freedman 1985), although \markcite{mea}Massey et al.~(1995) 
argue that NGC 6822 is deficient in both the most massive stars and the most 
populous OB associations compared to the Galaxy and the Magellanic Clouds 
(MCs).  

The optical structure of NGC 6822 has been studied by \markcite{h77}Hodge 
(1977) and \markcite{hea}Hodge et al.~(1991).  The system is dominated by a
strong bar oriented roughly north-south.  The brightest \hii\ regions are to 
the north of the bar. \markcite{md}Markert \& Donahue (1985) provide an
excellent image (their Fig.~1) showing the overall optical structure of the 
system.  The absolute magnitude of NGC 6822 is $M_B \approx -15.9$, just over a 
magnitude fainter than the SMC.  \markcite{hea}Hodge et al.~(1991) find that 
the density of resolved stars (from Schmidt plates), as well as the light from 
the unresolved stars can be fit with an exponential with a scale length of 
$\sim$120$''$.  They also find a limiting radius of $\sim$40$'$.  
\markcite{rc3}deVaucouleurs et al.~(1991, hereafter RC3) quote a major axis 
diameter of $D_{25} = 15'\llap.5$.

The \hi\ structure and content of NGC 6822 has been studied by 
\markcite{s87}Skillman (1987) and \markcite{gw}Gottesman \& Weliachew (1977).
NGC 6822 has $M_{H\,I}/L_B \approx 0.6$ (in Solar units), and its \hi\ emission 
is far more extended than its optical emission.  These are typical properties 
for a dI galaxy.  NGC 6822 is unusual in that the optical and \hi\ major axes 
are misaligned by almost 90$^\circ$ (see Fig.~11 in \markcite{hea}Hodge et 
al.~1991).  CO emission has been detected at a number of locations in NGC 6822 
(e.g.\markcite{wco}Wilson 1992b; \markcite{oea}Ohta et al.~1993; 
\markcite{w94}Wilson 1994; \markcite{itb}Israel, Tacconi \& Baas 1995).  In
general, the CO peaks are associated with either the brightest \hii\ regions, 
or with peaks in the \hi\ surface density.  NGC 6822 has a rich population of 
\hii\ regions, indicating substantial recent and on-going star formation.  The 
most complete study of the \hii\ regions of NGC 6822 is that of Hodge and 
collaborators (\markcite{chk}Collier, Hodge \& Kennicutt 1995, and references 
therein).  NGC 6822 is near enough that it is resolved in the {\it IRAS} 
far-infrared (FIR) data (\markcite{fir}Gallagher et al.~1991).  There is a 
general correspondance between the FIR emission and both the \hii\ regions and 
peaks in the \hi\ distribution, with some evidence for diffuse FIR emission as 
well.

The two main techniques for estimating heavy element abundances in Local Group
objects are spectrophotometry of nebular emission-line objects, and isochrone 
fitting onto color-magnitude diagrams (CMDs) of resolved stellar populations.  
For NGC 6822, studies of \hii\ regions (\markcite{stm}Skillman, Terlevich \& 
Melnick 1989) and planetary nebulae (\markcite{rmc}Richer \& McCall 1995) 
indicate similar oxygen abundances.  The \hii\ region result is $12+\log(O/H)
=8.20$, and the planetary nebulae result is $12+\log(O/H)=8.1 \pm 0.1$.  This 
corresponds to 0.15--0.20$Z_{\odot}$, adopting $12+\log(O/H)_{\odot} = 8.93$ 
(\markcite{ga}Grevesse \& Anders 1989).  The CMD study of \markcite{gea}Gallart 
et al.~(1996b) is the best available for isochrone fitting.  They conclude that
their data support $[Fe/H] \approx -1.0$ for the most metal-rich population.  
Given the systematic ambiguities of relating gas-phase oxygen abundances to 
mean metallicities from isochrone fitting to resolved stellar populations, 
these results are in reasonable agreement.

The x-ray populations of the Galaxy and the MCs have been extensively studied 
with all available x-ray telescopes over the last several decades 
(e.g.~\markcite{s96}Snowden 1996; \markcite{v93}Verbunt 1993; 
\markcite{bop}Bradt, Ohashi \& Pounds 1992; \markcite{ww}Wang \& Wu 1992; 
\markcite{wea}Wang et al.~1991).  There have also been studies of the x-ray 
source populations in M31 (e.g.~\markcite{s97}Supper et al.~1997; 
\markcite{pfj}Primini, Forman \& Jones 1993; \markcite{ftv}Fabbiano, Trinchieri 
\& van Speybroeck 1987) and M33 (e.g.~\markcite{lea}Long et al.~1996; 
\markcite{tfp}Trinchieri, Fabbiano, \& Peres 1988).  As noted above, there has 
been less work on the more optically faint galaxies in the Local Group.  The 
x-ray emission from M32 is dominated by an unresolved source that may be 
emission from low-mass x-ray binaries (LMXRBs) or a $\mu$AGN 
(\markcite{ewd}Eskridge et al.~1996).  The x-ray emission in the direction of 
the Local Group dI IC 1613 is dominated by a background galaxy cluster 
(\markcite{e95}Eskridge 1995).

The organization of this paper is as follows:  In \S 2 we review the existing
x-ray data for NGC 6822.  These data reveal a single bright point source, and
no evidence for diffuse emission.  In \S 3 we examine the optical information
on the environment of the x-ray point source.  We propose various 
interpretations for the nature of the source in \S 4.  In \S 5 we discuss these
possibilities, and outline the observations required to distinguish between 
them.  Finally, \S 6 presents a summary of our results.

\section{X-Ray Observations}

NGC 6822 was observed with the {\it Einstein} Observatory HRI 
(\markcite{ein}Giacconi et al.~1979) for 34.4 kiloseconds on 31 Oct.~1979 
(\markcite{md}Markert \& Donahue 1985).  This observation detected a bright 
point source in the direction of the bar, and another source that appears to be 
in the background (see Fig.~1 in \markcite{md}Markert \& Donahue 1985; see also
Fig.~7 of \markcite{p0}Fabbiano, Kim \& Trinchieri 1992).  \markcite{md}Markert 
\& Donahue (1985) point out that the source in the bar falls within the 
emission line object Ho 12 (\markcite{h77}Hodge 1977), and suggest that the 
x-ray source is related to this object.  The count-rate of the source from the 
{\it Einstein} HRI exposure is 0.0043 s$^{-1}$.  For a foreground column of 
$N_H = 9 \times 10^{20}~{\rm cm^{-2}}$ (\markcite{nh}Stark et al.~1992), and 
{\it assuming} a 5 keV bremmstrahlung spectrum (we shall return to this point
shortly), this corresponds to a flux of $8.56 \times 10^{-13}~{\rm 
erg~cm^{-2}~s^{-1}}$ in the 0.2--4.0 keV band (or a luminosity of $L_X \approx 
2.6 \times 10^{37}~{\rm erg~s^{-1}}$).  This is a perfectly reasonable 
luminosity for a LMXRB (e.g.~\markcite{wnp}White, Nagase \& Parmar 1995).  

The source was reobserved with the {\it ROSAT} PSPC in 1992 (sequence
\#RP600148).  Although the right ascensions from the HRI and PSPC observations 
agree to $\sim 1{''}\llap.5$, the declinations differ by $\sim$12$''$, with the
PSPC position north of the HRI position.  This is larger than the nominal 
positional accuracy of either instrument, and is thus a cause for some concern. 
We found that the second HRI source was also recovered in the PSPC observation, 
and showed the same offset in declination.  Thus there is a systematic offset 
in the aspect solutions for the two observations.  Neither source has any 
obvious optical or radio counterpart, however there are seven additional 
sources detected in the PSPC observation.  None of these additional sources 
are coincident with any cataloged radio source, or with any optical source down
to the sky survey limit.  However, we find that two of these sources (1RXP 
J194427$-$1443.0 and 1RXP J194505$-$1436.4) are $\sim$12$''$ north of 
moderately bright stars (once again, there are no cataloged radio 
counterparts).  From this conclude that the systematic error is in the PSPC 
astrometry for the field, perhaps due to the same problem discussed by 
\markcite{hpc}Harris, Perley \& Carilli (1994).

The PSPC exposure was quite short; the dead-time corrected exposure is 6364 s, 
resulting in $\sim$75 total source counts.  This corresponds to a PSPC 
count-rate of 0.0012s$^{-1}$.  We extracted the source spectrum from a 2$'$ 
radius circle, centered on the source.  We used a background spectrum drawn 
from an annular region, also centered on the source, extending from 2$'$ to 
$3'\llap.75$.  We used {\sl XSPEC v9.0} (\markcite{arn}Arnaud 1996) to assess a 
variety of spectral models.  However, due to the small number of source counts, 
the spectral parameters are rather poorly determined.  Statistically acceptable 
fits exist for both Raymond-Smith thermal (RS) and blackbody (BB) models (see 
Table 1).  The RS model yields acceptable fits for two distinct temperatures, 
with kT$\approx$0.04 keV for the soft solution, and kT$\approx$0.56 keV for the 
hard solution.  The BB model gives an acceptable fit for the soft solution only 
(with kT$\approx$0.06 keV).  As noted above, the observed {\it Galactic} column 
in this direction is $N_H = 9 \times 10^{20}~{\rm cm^{-2}}$.  The observed 
intrinsic column associated with this region of NGC 6822 is $N_H \approx 2 
\times 10^{21}~{\rm cm^{-2}}$ (see \markcite{hea}Hodge et al.~1991).  Thus we 
can constrain good fits to have $N_H$ in the range $\sim$1--3$\times$10$^{21} 
{\rm cm^{-2}}$.  This argues against the soft RS fit unless the source is 
strongly self-absorbed.  Both the hard RS fit and the BB fit have $N_H$ in the 
acceptable range.  In Figure 1, we show the observed spectrum, along with the 
fits for the 0.56 keV RS model and 0.06 keV BB model.  We show the $\chi^2$
confidence contours of $N_H$ and kT for these models in Figure 2.  The contours
are at 1$\sigma$, 90\%, and 99\% levels.  In particular, we note that a hot,
highly absorbed model is strongly ruled out by the data.

The spectral models discussed above yield a flux of only $\sim 8 \times 
10^{-14}~{\rm erg~cm^{-2}~s^{-1}}$ in the 0.2--2.0 keV band (this value changes 
by $\la$10\% when computed for the {\it Einstein} band).  This is an order of
magnitude lower than the published {\it Einstein} flux.  However the {\it 
Einstein} flux was computed for an assumed spectral model that is quite 
different from (and much harder than) those we find for the PSPC data.  We 
tested the importance of this difference by using the spectral models that fit 
the PSPC data to predict the {\it Einstein} HRI count-rate {\it a posteriori}.  
The results of this experiment are given in Table 1; for all good PSPC spectral
models, the predicted {\it Einstein} HRI count-rate is still roughly half a dex
lower than the observed {\it Einstein} HRI count-rate.  Thus the source is 
x-ray variable by at least half a dex on $\sim$decade timescales.  We searched 
for evidence of short-term variability in the PSPC data by inspecting the count 
rates for the individual observation intervals.  Any variability in the data 
can be easily explained by fluctuations in the background.  This is not a 
terribly strong constraint, given the small number of source counts.  

\section{Source Environment}

\subsection{Association with Ho 12}

As the source is superposed on the bar of NGC 6822, one might expect it to be 
an LMXRB.  However, as LMXRBs tend to have temperatures of 3--5 keV 
(e.g.~\markcite{wnp}White et al.~1995), the PSPC data argue strongly against
this interpretation.  As noted by \markcite{md}Markert \& Donahue (1985), the
source is projected on the emission-line object Ho 12:  $\Delta\alpha =
0{^s}\llap.2$, $\Delta\delta = 4''$, where we take the x-ray position from
\markcite{p0}Fabbiano et al.~(1992), and the Ho 12 region position from
\markcite{hkl}Hodge, Kennicutt \& Lee (1988).  In Table 2, we give these 
positions precessed to J2000 coordinates.  As noted in \S 2, above, this 
places the x-ray source within the boundaries of Ho 12.  Figure 3 is an 
$I$-band image centered on the x-ray source that was kindly provided by 
C.~Gallart.  The region shown is 54$''$ on a side.  The contours are H$\alpha$ 
emission from \markcite{ch}Collier \& Hodge (1994).  The circle centered on the 
x-ray position is 10$''$ in radius, and contains the stellar objects listed in 
Table 3.  These stars are identified from the photometry of 
\markcite{gav}Gallart et al.~(1996a).  Positions for the stars in Table 3 were 
computed from the astrometry of \markcite{w95}Wilson (1995) for nearby stars 
common to the \markcite{w92}Wilson (1992) and \markcite{gav}Gallart et 
al.~(1996a) studies.  Our positions are on \markcite{w95}Wilson's (1995) 
astrometric system to $\sim$1$''$.  We also give magnitudes and colors for the 
stars from the photometry of \markcite{gav}Gallart et al.~(1996a), and the 
offsets between the stellar positions and that of the x-ray source.  There are 
17 stars listed in Table 3.  Not all of these stars have measured $I$-band 
magnitudes, and so not all are visible on Figure 3.  A number are visible, but 
are too severely blended for accurate photometry.  

\subsection{The Nature of Ho 12}

Ho 12 was identified as an emission-line object by \markcite{h69}Hodge (1969;
\markcite{h77}1977), and is included in the study of the properties of the 
\hii\ regions of NGC 6822 by him and his collaborators (\markcite{chk}Collier 
et al.~1995).  It is, however, unclear if this object is an \hii\ region or a 
supernova remnant (SNR).  \markcite{s75}Smith (1975) obtained an IDS spectrum 
of Ho 12 with a wavelength coverage of $\sim$2400\AA, from H$\beta$ to the 
[\sii] $\lambda\lambda$6717,6731\AA\ doublet (he uses the 
\markcite{h69}Hodge 1969 nomenclature, and thus refers to the object as \#10.  
This is the same object as Ho 12, following \markcite{h77}Hodge 1977).  He did 
not publish his spectrum, however he states that the [\sii] doublet ratio 
implies a high electron density, that the [\sii] to H$\alpha$ ratio is quite 
high ($\sim$0.5) and that [\oi] $\lambda$6300\AA~is observed.  All of these 
argue against Ho 12 being a typical \hii\ region.  

\markcite{ddb}D'Odorico, Dopita \& Benvenuti (1980) obtained narrow-band 
photographic images of Ho 12 in H$\alpha$ and [\sii] with the 48$''$ Schmidt
telescope at Palomar Observatory.  Considering both their imaging and 
\markcite{s75}Smith's spectrophotometry, they argue that Ho 12 is actually an 
SNR.  They state that subsequent AAT spectroscopy confirm this designation, but 
these data have never been published.  The case for Ho 12 being an SNR is 
reasonable, but it would be very useful to obtain new spectrophotometry.  

\markcite{w92}Wilson (1992a) identifies an OB association (her OB11) containing 
Ho 12 (\markcite{h77}Hodge's (1977) OB10 is a subset of her OB11).  Based on 
her CM diagram (shown in her Fig.~6), she estimates an age of 
$\sim$8$\times$10$^6$ yr for OB11.  There are stars with $(B-V) \approx 0$ as 
bright as $V \approx 18$ in her CMD.  These stars have absolute magnitudes of 
$M_V \approx -5.5$, roughly commensurate with O5 stars.  There are thus clearly 
stars hot and luminous enough to support an \hii\ region as bright as Ho 12.  
There is also a rich population of red supergiants (reaching $V \approx 18$, 
$(B-V) \approx 2$), demonstrating that the star-forming event began 
sufficiently long ago for the most massive stars to have already gone 
supernova.  Thus the stellar population in the vicinity does not offer any firm 
guidance for deciding if Ho 12 is an \hii\ region or an SNR.  We will therefore 
consider both interpretations in our analysis below.

\section{Possible Interpretations}

The source is {\it far} too cool to be a typical high-mass x-ray binary 
(HMXRB), as such sources typically have kT$\approx$10 keV 
(e.g.~\markcite{wnp}White et al.~1995).  SNR can have the sort of x-ray 
spectral properties observed (e.g.~\markcite{sss}Seward, Schmidt \& 
Slane~1995).  The evidence that Ho 12 is an SNR makes this a possibility well
worth considering.  There is no radio continuum emission from the source down 
to limits of 25 mJy at 21 cm (\markcite{gw}Gottesman \& Weliachew 1977), 42 mJy
at 6cm (\markcite{gea}Griffith et al.~1994), and 2.7 mJy at 2.8cm 
(\markcite{kgw}Klein, Gr\"ave \& Wielebinski 1983).  This is probably not a 
strong constraint:  \markcite{m83}Mathewson et al.~(1983, \markcite{m85}1985)
present radio continuum and x-ray data for a sample of MC SNR.  We use their
data to define a typical range in x-ray flux to radio power for SNR.  This 
range is $\sim$2 dex, thus it has only rough predictive power.
\markcite{m83}Mathewson et al.~(1983, \markcite{m85}1985) give radio data at 
either 408 or 843 MHz.  We correct this to the observed frequencies for NGC 
6822 by taking a typical SNR radio spectral index of $\alpha \approx -0.5$.  
The result of this excercise is that the {\it expected} radio power of a SNR
with the observed $f_X$ is on the order of the current observational limits, if 
it has properties like those of the SNR in the MCs.  

A more compelling argument against the x-rays being from an SNR is the x-ray 
variability between the {\it Einstein} and {\it ROSAT} observations.  Of 
course, this assumes that the x-rays observed by both {\it Einstein} and {\it 
ROSAT} come from the same object.  It is possible that the {\it Einstein} data 
are dominated by emission from a highly variable compact x-ray source that was
quiescent during the PSPC observation, and that the PSPC data are entirely due
to emission from Ho 12.  The H$\alpha$ emission from Ho 12 is clearly extended 
(see Figure 3); the main structure has a diameter of $\sim$10$''$.  This is 
substantially less than the point-spread function of the {\it ROSAT} PSPC, thus 
we would not expect the PSPC emission to be resolved (nor do we observe it to 
be) even if it were due to emission from Ho 12.  The {\it Einstein} HRI 
emission is unresolved, but a small enough fraction of the HRI counts would be 
due to the extended source that this is not a useful constraint.  Although such 
a multi-source model is unappealing when viewed from the edge of Occam's razor, 
the available data do not rule it out.  One way of testing this two-source
picture would be to obtain a deep {\it ROSAT} HRI image to search for evidence
of extended emission.  Keeping this in mind, we turn our attention to 
possible interpretations for the emission assuming that only one source of
x-rays is present.

\subsection{Is the Object a Super-Soft Source?}

We suggest that this source may be a ``Super-Soft'' source (SSS), such as were
originally discovered in the MCs (\markcite{lhg}Long, Helfand \& Grabelsky 
1981).  X-ray emission from these sources is best fit by optically thick 
(blackbody) models, with temperatures of $\sim$50 eV, and x-ray luminosities up 
to $\sim$10$^{38}~{\rm erg~s^{-1}}$ in the {\it ROSAT} band.  The PSPC data for 
the NGC 6822 source are entirely consistent with these spectral properties.  
The luminosity of the source is a bit low in the PSPC observation, but the 
source was $\sim$10 times brighter in the {\it Einstein} HRI observation; the 
maximum observed luminosity ($L_X \approx 2.6 \times 10^{37}~{\rm erg~s^{-1}}$) 
is in keeping with observations of other SSSs.  Also, recent observations of
CAL 83 discussed in \markcite{aea}Alcock et al.~(1997) have shown it to exhibit 
strong x-ray variability.  This may be a typical property of SSSs.

The nature of SSSs is still unclear.  There are several currently debated
models for SSSs, and it is likely that they are not a homogenous class
(e.g.~\markcite{kph}Kahabka, Pietsch \& Hasinger 1994).  One idea is that SSSs 
are due to high accretion-rate nuclear burning on white dwarfs 
(e.g.~\markcite{k95}Kahabka 1995).  This model is particularly well developed 
for the transient SSSs.  Another proposed mechanism is emission from the hot 
inner part of white dwarf accretion disks.  It has also been proposed that SSSs 
are due to accretion onto black holes, or onto neutron stars.  Especially in 
the case of neutron stars, this requires the energy source to be enveloped in 
an optically thick cocoon of material in order to produce x-rays as soft as 
those observed.  See \markcite{cea}Crampton et al.~(1996) and 
\markcite{kph}Kahabka et al.~(1994) for further discussion of these models, and 
references to earlier work.

\subsection{Is the Object a Stellar-mass Black Hole Binary?}

The spectrum can also be fit with a model that is substantially hotter than any
appropriate for a SSS.  The hotter fit has spectral properties similar to
those observed for a number Galactic and MC stellar-mass black
hole binary candidates (e.g.~\markcite{c92}Cowley 1992 and references therein). 
Thus, if the hotter fit is correct, the best working hypothesis for the source 
is that it is a black hole binary system.  The x-ray spectral properties of 
black-hole binary candidates are heterogenous ({\it viz}.~\markcite{c92}Cowley 
1992; \markcite{mhr}McClintock, Horne \& Remillard 1995).  While they typically 
have fairly soft spectra (kT$\la$1.5 keV; indeed, the SSS CAL 87 is a good 
black-hole binary candidate, as discussed by \markcite{c92}Cowley 1992), some 
sources exhibit substantial x-ray spectral variability, and can have states 
with spectra characterized by temperatures of up to kT$\approx$30 keV 
(e.g.~V616 Mon).

\section{Discussion}

\subsection{Optical Identification and Possible Follow-up Observations}

In \S 3.1 we presented information from the photometry of \markcite{gav}Gallart 
et al.~(1996a) on 17 stars within 10$''$ of the x-ray position.   Observations 
of Galactic and MC SSSs and black-hole binary candidates show them to typically 
have $-2 \la M_V \la +0.5$ and $(B-V) \approx 0$ (e.g.~\markcite{cea}Crampton 
et al.~1996; \markcite{bea}Beuermann et al.~1995; \markcite{c93}Cowley et 
al.~1993; \markcite{c92}Cowley 1992 and references therein).  For an object at 
the distance of NGC 6822, the above apparent magnitude range corresponds to 
$21.5 \la m_V \la 24.0$.  The faintest objects in Table 3 have $m_B \approx 
22.5$, thus it is entirely possible that the optical counterpart to the x-ray 
source is undetected in the existing photometry.  Among the stars listed in 
Table 3, the most likely candidates for the optical counterpart are those with 
the bluest colors, and within 5$''$ of the x-ray position:  Stars 17801, 17896, 
17996, 18260, and 18332.  We are seeking to obtain follow-up spectroscopy of 
these candidates.  While this is clearly warrented, deeper photometry from HST 
is likely to be required to find the optical counterpart.  Work on the UV 
properties of Galactic and MC SSSs and black-hole binary candidates 
(e.g.~\markcite{tea}Treves et al.~1990; \markcite{bea}Beuermann et al.~1995) 
argues that the object will have an unabsorbed near-UV flux in the range 
$10^{-16}$--$10^{-18}~{\rm erg~cm^{-2}~s^{-1}~\AA^{-1}}$.  For $E(B-V) \approx 
0.36$ \markcite{m83}McAlary et al.~(1983), a STIS NUV image of the region 
centered on the x-ray position should detect the source at the 
10$\sigma$-level, even in it's faint state, in an integration time of 
$\sim$2000 seconds.  The OB 11 turn-off stars in the vicinity have $V \approx 
21$.  These stars should also be easily detected in such an NUV image, making 
it possible to tie the astrometry between the optical and UV observations, and 
unambiguously identify the UV sources in the field.  If the source is UV-bright 
($\sim 10^{-16}~{\rm erg~cm^{-2}~s^{-1}~\AA^{-1}}$) it is likely to be a 
black-hole binary.  If it is closer to the faint end of the range, this would 
support the SSS interpretation.  We plan on proposing for such observations in 
order to study the properties of the source in more detail.

It would be interesting to obtain better x-ray data on the NGC 6822 source.
As noted in \S 4, a deep {\it ROSAT} HRI observation could test the
possibility that there are two sources of x-rays, one being faint extended 
non-variable emission from Ho 12, and the other being compact highly variable 
emission from an x-ray binary.  {\it ASCA}, {\it RXTE}, or {\it SAX} 
observations could, in principle, provide much better spectral constraints than 
the PSPC data.  However, the soft spectrum poses something of a problem, 
especially if the source happens to be in a low-flux state when observed.  As 
an illustrative example, if the source were observed for 20 ksec by {\it ASCA}, 
while in its low-flux state, the resulting spectrum would provide no 
improvement over the $\sim$6 ksec PSPC spectrum.  Similar arguments hold for 
{\it RXTE} and {\it SAX}.  It is unlikely that improved x-ray spectral data for 
the NGC 6822 source will be obtained until {\it AXAF} is in orbit.

\subsection{Comparison of NGC 6822 with Other Local Group Galaxies}

It is useful to compare the x-ray properties of NGC 6822 with those of the MCs
and M31.  The MCs are low-luminosity star-forming systems much like NGC 6822, 
but are somewhat more luminous, and a great deal closer.  Their x-ray 
properties have therefore been studied in some detail.  M31 is the closest 
giant galaxy, and has also been well studied in the x-ray.

The x-ray emission in NGC 6822 and the MCs is mainly due to a very small number 
of discrete sources.  Thus we compare the source populations of the three
systems.  In order to do so, we adopt the total blue magnitudes for all three 
galaxies as given in the \markcite{rc3}RC3 (${B_T^0} = 0.57$, 2.28, and 8.39 
for the LMC, the SMC, and NGC 6822 respectively), and distances of 50 kpc for 
the LMC, and 65 kpc for the SMC (as noted in \S 1, the distance to NGC 6822 is 
$\sim$500 kpc).  \markcite{sea}Schmidtke et al.~(1994) and \markcite{mcs}Cowley 
et al.~(1997) present the current results for the x-ray source populations of 
the MCs.  These studies are based on {\it ROSAT} HRI observations, and thus do 
not provide spectral information.  If we scale our observed PSPC count-rate to 
an expected HRI count-rate for models appropriate for our source, and then 
correct these count-rates to the MC distances, we find that the source in NGC 
6822 would have an HRI count-rate of $\sim$0.040 ${\rm s^{-1}}$ were it in the 
LMC, or $\sim$0.023 ${\rm s^{-1}}$ were it in the SMC.  \markcite{sea}Schmidtke 
et al.~(1994) and \markcite{mcs}Cowley et al.~(1997) indentify half a dozen 
sources in each galaxy that are at or above these limits.  From the apparent
magnitudes and distances given above, the expected numbers are $\sim$13 in the 
LMC, and $\sim$5 in the SMC.  Given the small numbers involved, and the
crudeness of the comparison, this is a surprisingly good result.

For M31, we adopt a total blue magnitude of ${B_T^0}=3.34$ from the 
\markcite{rc3}RC3, and a distance of 690 kpc.  \markcite{s97}Supper et 
al.~(1997) find $\sim$200 sources in M31 that are as luminous as the NGC 6822 
source.  As the optical luminosity of M31 is $\sim$200 times that of NGC 6822, 
this comparison suggests that the number of discrete x-ray sources does scale 
quite well with optical luminosity.  

The issue of how galaxy morphology affects x-ray source population is still
quite mysterious.  The above arguments indicate that low luminosity galaxies
are just as efficient at producing stellar x-ray sources as high luminosity
galaxies.  However, their are other indications that this is not so.  {\it
ROSAT} observation of the Fornax dwarf spheroidal (\markcite{gmd}Gizis, Mould
\& Djorgovski 1993) found no evidence of any stellar x-ray sources associated 
with Fornax.  Comparing Fornax's optical luminosity to those of Galactic
globular clusters, \markcite{gmd}Gizis et al.~(1993) concluded that the
extremely low density environment of Fornax is significantly less efficient
at producing stellar x-ray sources than are environments such as globular
clusters.  The problem with all of this is the small number of sources found
associated with the small number of galaxies that have been studied.  It would
be very interesting to determine how the x-ray source populations of diffuse
dwarf galaxies (both early- and late-types) differ from those of more
luminous systems.  We certainly cannot draw any firm conclusions in this
paper, but it may be possible to address this issue sensibly by studying the
composite x-ray properties of a large sample of dwarf galaxies.

\section{Summary}

A bright x-ray point source in the bar of NGC 6822 was originally discovered 
in an {\it Einstein} HRI image (\markcite{md}Markert \& Donahue 1985).  A
subsequent {\it ROSAT} PSPC observation reveals that the source is variable
by roughly a dex on the timescale of a decade, and has has a very soft x-ray 
spectrum.  The PSPC observation was quite short ($\sim$6 ksec), resulting in 
only $\sim$75 source counts.  Thus the spectral properties are not well 
constrained.  We can fit the data with either a RS model with kT$\approx$ 0.56 
keV, or with a BB model with kT$\approx$0.06 keV.  The spectral properties of 
the source resemble those of Galactic and MC SSSs or black hole binary 
candidates.  

The source is superposed on the emission-line object Ho 12.  Existing data on
this object argue that it may be either an \hii\ region or a SNR.  The 
variability of the x-ray source indicates that the x-rays are not due to 
direct emission from an SNR.  We have identified a list of potential optical 
counterparts from the photometry of \markcite{gav}Gallart et al.~(1996a), and
are seeking to obtain follow-up spectroscopy of these objects.  However, 
comparison with optical data for similar objects in the Galaxy and the MCs 
indicates that the source may be too faint to appear in existing optical data.  
Optical and UV imaging from HST is the best currently available means of 
searching for a counterpart.

\acknowledgments

We are happy to thank Carme Gallart for access to her photometry and $I$-band
image of NGC 6822, and Paul Hodge for allowing us to use his H$\alpha$ data.
Jack Hughes gave us advice on the x-ray properties of SNR, and Dan Harris on 
x-ray astrometry.  Martin Elvis suggested that we include the $\chi^2$ 
confidence plots.  Harley Thronson was the PI on the PSPC observing proposal 
for NGC 6822.  We thank the referee, Niel Brandt, for a number of very useful
suggestions.  This research has made use of the NASA/IPAC Extragalactic 
Database (NED) which is operated by the Jet Propulsion Laboratory, Caltech, 
under contract with the National Aeronautics and Space Administration.  This 
research was partially supported by the National Aeronautics and Space
Administration under ROSAT Grant No.~NAG 5-1973.

\newpage

{

\def\tabrule{\noalign{\hrule}}
\def\pz{\phantom{0}}
\def\pb{\phantom{-}}
\def\pd{\phantom{.}}
\

\centerline{Table 1 -- PSPC Model Fitting Results}
\vskip0.3cm

\newbox\tablebox
\setbox\tablebox = \vbox {

\halign{#&\hfil\pz#\hfil&\hfil\pz#\hfil&\hfil\pz#\hfil&\hfil\pz#\hfil&\hfil#
\hfil&\hfil#\hfil&\hfil\pz#\hfil&\hfil\pz#\hfil\cr
\tabrule
\noalign{\vskip0.1cm}
\tabrule
\noalign{\vskip0.1cm}

Model & kT & 90\% conf. & $N_H$ & 90\% conf. & ${f_X}^1$ & ${f_X}^1$ & 
${\rm HRI}_{Ein}^2$ & $\chi^2/\nu$ \cr
\ & keV & & $10^{21}~{\rm cm^{-2}}$ & & 0.2--2 keV & 0.2--4 keV & $10^{-3}~{\rm
s^{-1}}$ & \cr
\noalign{\vskip0.1cm}
\tabrule
\noalign{\vskip0.2cm}
R-S$_{hot}$ & 0.56$\pz$ & 0.090--0.79 & 0.39 & $\pz$0.1--19.7 & 8.25 & 8.38 & 
1.30 & 16.8/17=0.99 \cr
R-S$_{cool}$ & 0.044 & 0.042--0.79 & 22.3 & 18.9--27.6 & 7.05 & 8.23 & 1.49 & 
17.8/17=1.05 \cr
Blackbody & 0.064 & 0.029--0.14 & 12.7 & $\pz$2.1--32.8 & 7.28 & 7.84 & 1.47 & 
17.4/17=1.03 \cr
\noalign{\vskip0.2cm}
\tabrule
}
}
\centerline{ \box\tablebox}

1:  Fluxes are given in units of 10$^{-14}~{\rm erg~cm^{-2}~s^{-1}}$.

2:  Predicted count-rate for the {\it Einstein} HRI.  The observed {\it 
Einstein} HRI count-rate is $4.21 \times 10^{-3}~{\rm sec^{-1}}$ 
(\markcite{p0}Fabbiano et al.~1992).

Abundances for all fits were fixed at 0.25 Solar.

}

\vskip100pt

{

\def\tabrule{\noalign{\hrule}}
\def\pz{\phantom{0}}
\def\pb{\phantom{-}}
\def\pd{\phantom{.}}
\

\centerline{Table 2 -- X-Ray and Optical Emission-Line Source Positions} 
\vskip0.3cm

\newbox\tablebox
\setbox\tablebox = \vbox {

\halign{\pz#\pz\hfil&\hfil\pz#\pz\hfil&\hfil\pz#\pz\hfil\cr
\tabrule
\noalign{\vskip0.1cm}
\tabrule
\noalign{\vskip0.1cm}

\ & RA(J2000.0) & Dec(J2000.0) \cr
\ & h~m~s & $\circ$ $'$ $''$ \cr
\noalign{\vskip0.1cm}
\tabrule
\noalign{\vskip0.2cm}
Ho 12 & 19 44 56.9 & $-$14 48 27 \cr
X-Ray & 19 44 56.7 & $-$14 48 31 \cr
\noalign{\vskip0.2cm}
\tabrule
}
}
\centerline{ \box\tablebox}

}

\newpage

{

\def\tabrule{\noalign{\hrule}}
\def\pz{\phantom{0}}
\def\pb{\phantom{-}}
\def\pd{\phantom{.}}

\centerline{Table 3 -- Data on Potential Optical Counterparts}
\vskip0.2cm

\newbox\tablebox
\setbox\tablebox = \vbox {

\halign{\pz#\hfil&\hfil\pz#\hfil&\hfil#\hfil&\hfil#\pz\hfil&\pz#\hfil
&\hfil\pz#\hfil&\hfil\pz#\hfil&\hfil#\pz\hfil&\hfil
#\pz\hfil\cr
\tabrule
\noalign{\vskip0.1cm}
\tabrule
\noalign{\vskip0.1cm}

ID & RA(J2000.0) & Dec(J2000.0) & offset & \hfil$B$ & $(U-B)$ & $(B-V)$ 
& $(V-R)$ & $(V-I)$ \cr
\ & h~m~s & $\circ$ $'$ $''$ & $''$ \cr
\noalign{\vskip0.1cm}
\tabrule
\noalign{\vskip0.2cm}
17594 & 19 44 57.2 & $-$14 48 29 & 7.1 & 20.57$\pm$0.03 & $\pb$0.66$\pm$0.09 & & & \cr
17633 & 19 44 57.1 & $-$14 48 25 & 8.4 & 21.61$\pm$0.07$^a$ & & & & 1.21$\pm$0.11 \cr
17683$^b$ & 19 44 57.1 & $-$14 48 37 & 8.2 & 22.19$\pm$0.07 & $-$0.26$\pm$0.14 & & & \cr
17801$^c$ & 19 44 57.0 & $-$14 48 33 & 4.3 & 21.68$\pm$0.09 & $-$0.82$\pm$0.12 & & & \cr
17896 & 19 44 56.9 & $-$14 48 31 & 2.3 & 21.44$\pm$0.06 & & & & \cr
17913 & 19 44 56.9 & $-$14 48 22 & 9.5 & 22.52$\pm$0.11 & $-$0.53$\pm$0.16 & & & \cr
17996 & 19 44 56.8 & $-$14 48 29 & 2.0 & 21.60$\pm$0.07 & & & & \cr
18013$^d$ & 19 44 56.7 & $-$14 48 26 & 5.1 & 21.58$\pm$0.13 & & & & \cr
18040$^e$ & 19 44 56.7 & $-$14 48 41 & 9.5 & 21.09$\pm$0.05 & $-$0.23$\pm$0.08 & & & \cr
18075 & 19 44 56.7 & $-$14 48 24 & 7.5 & 22.59$\pm$0.12 & $-$0.36$\pm$0.19 & 0.15$\pm$0.15 & $-$0.17$\pm$0.17 & 0.48$\pm$0.16 \cr
18194 & 19 44 56.5 & $-$14 48 39 & 8.0 & 21.57$\pm$0.07$^a$ & & & & \cr
18260 & 19 44 56.5 & $-$14 48 29 & 3.9 & 21.53$\pm$0.08 & $-$0.10$\pm$0.17 & 0.11$\pm$0.11 &  & 0.38$\pm$0.11 \cr
18332 & 19 44 56.4 & $-$14 48 32 & 4.5 & 22.50$\pm$0.10 & $-$0.56$\pm$0.17 & & & \cr
18435 & 19 44 56.3 & $-$14 48 39 & 9.6 & 20.67$\pm$0.14$^f$ & & & & \cr
18467 & 19 44 56.3 & $-$14 48 28 & 6.8 & 20.16$\pm$0.03 & $-$0.77$\pm$0.04 & & & \cr
18491 & 19 44 56.3 & $-$14 48 24 & 9.4 & 20.71$\pm$0.02 & $\pb$0.10$\pm$0.08 & 0.47$\pm$0.04 & $\pb$0.39$\pz$0.05 & 0.96$\pm$0.08 \cr
18630 & 19 44 56.1 & $-$14 48 35 & 9.8 & 20.70$\pm$0.02 & $\pb$1.19$\pm$0.13 & 1.31$\pm$0.04 & $\pb$0.89$\pz$0.05 & 1.74$\pm$0.06 \cr
\noalign{\vskip0.1cm}
\tabrule
}
}
\centerline{ \box\tablebox}

{\noindent{~$^a$ Magnitude is $V$-band.}\hfill\break
{~$^b$ $(B-R)$=0.77$\pm$0.09, $(B-I)$=1.36$\pm$0.12}\hfill\break
{~$^c$ $(B-I)$=1.73$\pm$0.11}\hfill\break
{~$^d$ $(B-I)$=1.40$\pm$0.17}\hfill\break
{~$^e$ $(B-I)$=1.76$\pm$0.07}\hfill\break
{~$^f$ Magnitude is $I$-band.}\hfill}
}

\newpage

\figcaption{Top panel:  The PSPC spectrum of the x-ray source (points with 
error bars), along with the best-fit warm (kT$\approx$0.56 keV) RS model 
(solid line), and the best-fit (kT$\approx$0.06 keV) BB model (dotted 
line).  Bottom panel:  $\chi^2$ residuals of the fits to the data.
}

\figcaption{$\chi^2$ confidence contours of kT and $N_H$ for a) the 0.56 keV RS 
model and b) the 0.06 keV BB model.  The best-fit values are shown as plus
signs.  The contours are at the 1$\sigma$, 90\%, and 99\% confidence levels.
}

\figcaption{An $I$-band image of the vicinity of the x-ray source.  North is 
up, East to the left.  The center of the field is at the x-ray position given 
in Table 2.  A 10$''$ radius circle is plotted, containing the stellar objects
listed in Table 3.  The contours are H$\alpha$ emission from Collier \& Hodge
(1994) for the optical emission line objects Ho 12 and HK 65. 
}

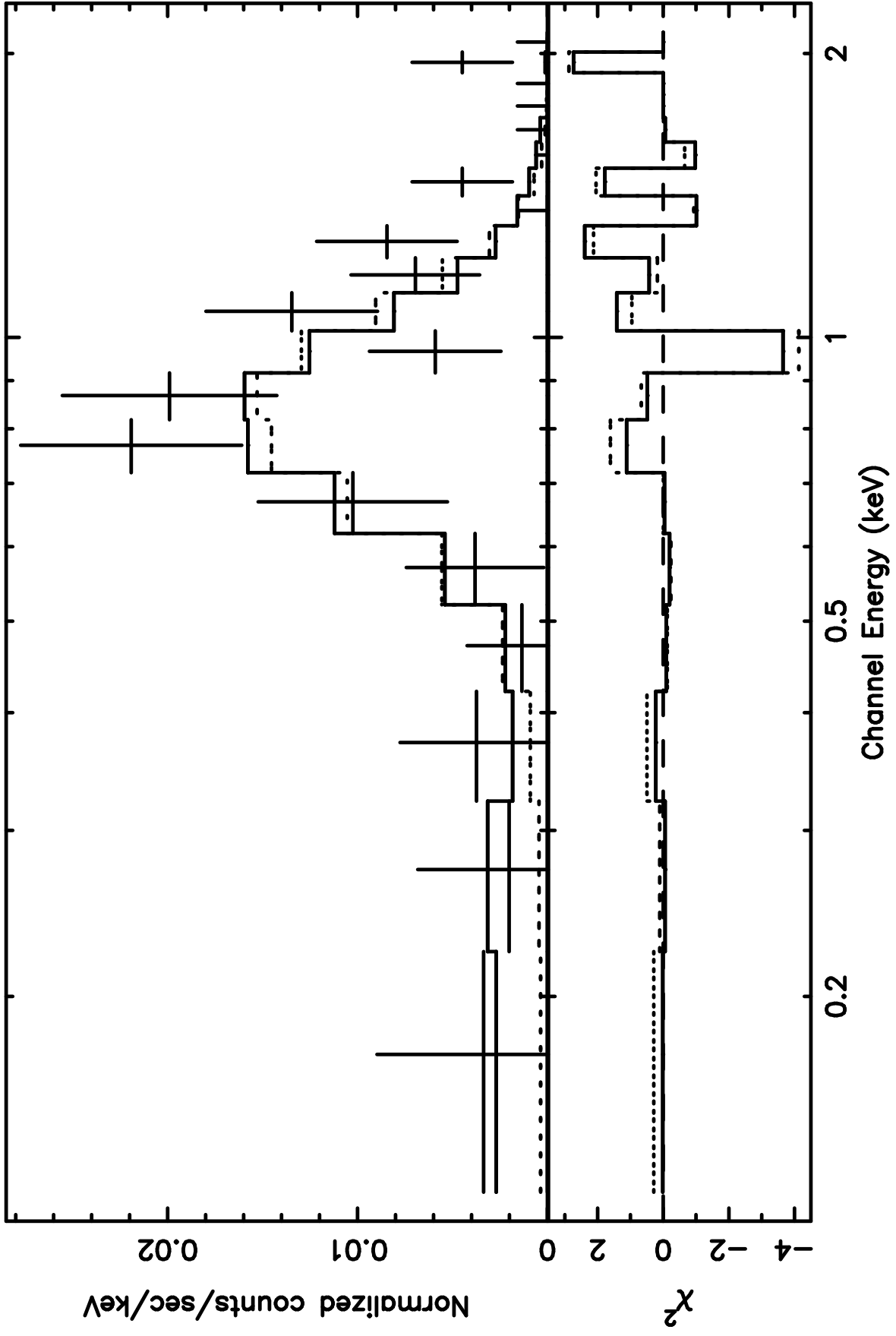
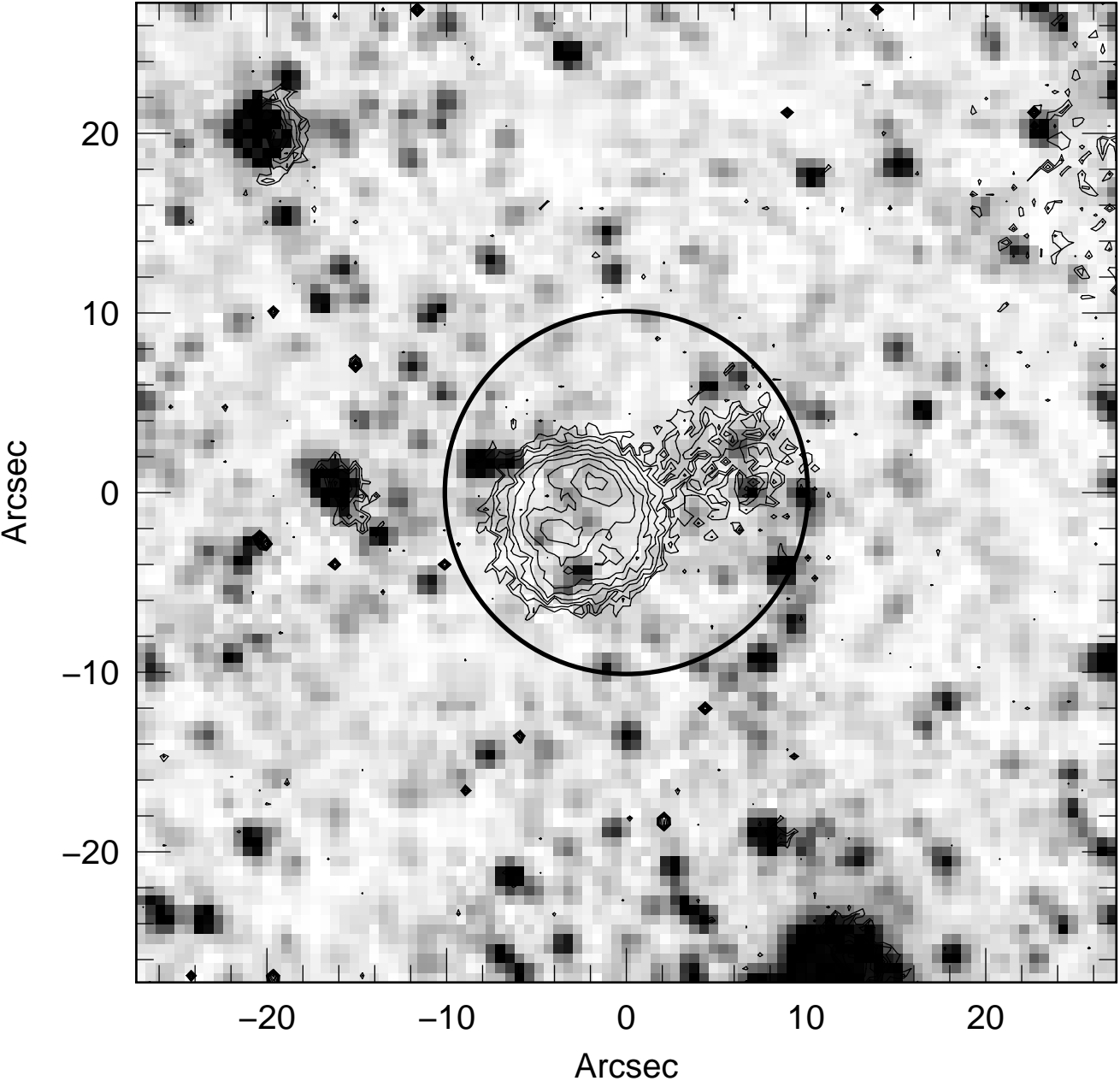
\end{document}